\documentclass[prl, twocolumn, superscriptaddress, showpacs]{revtex4}

\usepackage{graphicx} \usepackage{amssymb} \usepackage[usenames]{color}
\usepackage{amsmath} \usepackage{xspace}

\def\LCMO{$\mathrm{La_{0.50}Ca_{0.50}MnO_3}$\xspace}
\def\LCMOx{{La$_{1-x}$Ca$_{x}$MnO$_3$}\xspace}

\def\LCMOfiftwo{$\mathrm{La_{0.48}Ca_{0.52}MnO_3}$\xspace}
\def\PCMOfiftwo{$\mathrm{Pr_{0.48}Ca_{0.52}MnO_3}$\xspace}

\begin{document}

\title{Dirty Peierls transition to stripe phase in manganites} \author{S. Cox}
\affiliation{National High Magnetic Field Laboratory, Los Alamos National
Laboratory, Ms-E536, Los Alamos, New Mexico, 87545, USA}
\affiliation{Department of Materials Science and Metallurgy, University of
Cambridge, Cambridge, CB2 3QZ, UK} \author{J.C. Lashley} \affiliation{Los
Alamos National Laboratory, Los Alamos, New Mexico, 87545, USA} \author{E.
Rosten} \affiliation{Los Alamos National Laboratory, Los Alamos, New Mexico,
87545, USA} \author{J. Singleton} \affiliation{National High Magnetic Field
Laboratory, Los Alamos National Laboratory, Ms-E536, Los Alamos, New Mexico,
87545, USA} \author{A.J. Williams} \affiliation{Centre for Science at Extreme
Conditions, University of Edinburgh, Edinburgh, EH9 3JZ, UK} \author{P.B.
Littlewood} \affiliation{Cavendish Laboratory, University of Cambridge,
Cambridge, CB3 0HE, UK} \begin{abstract} The nature of the phase transitions in
\LCMOx and \PCMOfiftwo has been probed using heat capacity and magnetisation
measurements.  The phase transition associated with the onset of the stripe
phase has been identified as second order.  The model of a Peierls transition
in a disordered system (a `dirty' Peierls transition) is shown to provide an
extremely good fit to this transition.  In addition, an unexpected magnetic
phase has been revealed in low temperature \PCMOfiftwo, associated with an
excess heat capacity over a wide temperature range compared to \LCMOfiftwo.
\end{abstract}

\pacs{75.47.Lx, 71.45.Lr, 65.40.Ba, 65.40.Gr}

\maketitle

Many strongly correlated electron systems (e.g. manganites~\cite{CO1, CO2a},
cuprates~\cite{cuprates}, nickelates~\cite{nickelates} and
cobaltites~\cite{cobaltites}) exhibit charge ordering phenomena, in which a
superstructure forms at low temperatures.  The insulating nature of the
compounds and the results of transmission electron microscopy (TEM)
experiments~\cite{CO1, CO2a, chen_comm_incomm} led to the suggestion that the
superstructure formation was driven by charge separation and localisation at
atomic sites.  However, recent work has produced conflicting evidence as to the
nature of the superstructure, with some studies supporting a model with charge
localised at the atomic sites, but with the difference in charge between atomic
sites being small~\cite{rodcar, Daoud, proffen, PCMO_neutron, mnkedge1}, and
others indicating the the superstructure is not tied to the atomic
sites~\cite{us, pcmo_TEM}.  To explain the latter results, it has been proposed
that the superstructure resembles a charge density wave (CDW)~\cite{us,
milward}.  In this paper, we find strong support for a CDW model of the
superstructure in manganites.

Previous measurements of \LCMOx with  $x \geq 0.5$~\cite{Ramirez_therm_TEM,
Ramirez_cheong_schiffer, diaz2/3, Lees} have observed two transitions as peaks
in the heat capacity.  The peak at higher temperature ($T$) was attributed to
critical fluctuations of the order-disorder type associated with charge
ordering~\cite{Ramirez_therm_TEM, Ramirez_cheong_schiffer}, with a contribution
at $x=0.5$ from the onset of ferromagnetism (FM).  The transition was
identified as first order based on the hysteresis in the resistivity
data~\cite{diaz2/3}.  The lower $T$ peak was attributed to the transition from
a paramagnetic state to an antiferromagnetic (AFM)
state~\cite{Ramirez_therm_TEM, Ramirez_cheong_schiffer,  diaz2/3, Lees}.  

Here we use heat capacity and magnetisation measurements to gain insight into
the nature of the phase transitions in manganites.  \LCMO, \LCMOfiftwo and
\PCMOfiftwo were measured, with the latter two being chosen as compounds with
different average cation sizes and variances (see
Table~\ref{average_and_variance}) but in which the superstructure has on
average an almost identical wavevector~\cite{pcmo_TEM}.  The smaller size of
the Pr cation is thought to lead to stronger electron-phonon coupling, allowing
the superstructure to lock into the lattice in around 25\% of the
grains~\cite{pcmo_TEM}.  The \LCMO sample was chosen as it has a nominally
commensurate superstructure (though small deviations are seen in TEM
measurements~\cite{philmag}) and so provides a contrast between commensurate
and incommensurate systems.

\begin{table} \begin{tabular}{|l|c|c|} \hline & Average Re/Ae   &   Variance of
Re/Ae \\ & site radius (\AA)    & site radius (\AA$^2$)\\ \hline \LCMO
&                   1.198                                     &        $3.24
\times 10^{-4}$        \\    La$_{0.48}$Ca$_{0.52}$MnO$_3$ &     1.197
&        $3.23 \times 10^{-4}$        \\ Pr$_{0.48}$Ca$_{0.52}$MnO$_3$ &
1.180                                     &        $2.50 \times 10^{-7}$
\\ \hline \end{tabular} \caption{Average and variance of the radius of the site
occupied by rare earth (Re) or alkaline earth (Ae) ions in different
compositions of Re$_{1-x}$Ae$_x$MnO3 (the Re/Ae site radius). Here Re is La or
Pr, Ae is Ca and $x$=0.5 or 0.52.~\cite{shannon}.\label{average_and_variance}}
\end{table}

Samples were prepared  by repeated grinding, pressing and sintering of
appropriate oxides and carbonates in stoichiometric proportions.  The
carbonates were decarboxylated by heating for 12 hours at 950$^\circ$C.  Each
sample was reground, repelleted and heated at 1350$^\circ$C for 4 days, then
reground, repelleted and reheated at 1350$^\circ$C for another 4 days.  X-ray
powder diffraction indicated that the samples were single
phase~\cite{Williams}.

Heat capacity measurements were made using a Quantum Design Physical Properties
Measurement System (PPMS).  The accuracy of the measurements can be checked by
examining the fits to the PPMS-measured decay curves.  This is especially
important in the region of a first order transition, where the release of
latent heat can reduce the quality of the fit~\cite{PPMS}.  In order to ensure
than the system had reached equilibrium the heat capacity measurements were
taken with very dense data points (between 140 and 600 measurements made
between 1.8~K and 300~K), and the system was allowed around twenty minutes to
reach equilibrium at each $T$ (decreasing the waiting period to three minutes
produced substantially different data in the regions of the transitions).  The
long relaxation times at each $T$ hints at pinning of the superstructure to
defects in the system~\cite{gruner}.  Magnetic susceptibility measurements were
performed using a Quantum Design Magnetic Properties Measurement System.
Samples with masses between 30 and 45~mg were used.

\begin{figure} \begin{centering}
\includegraphics[width=0.47\textwidth]{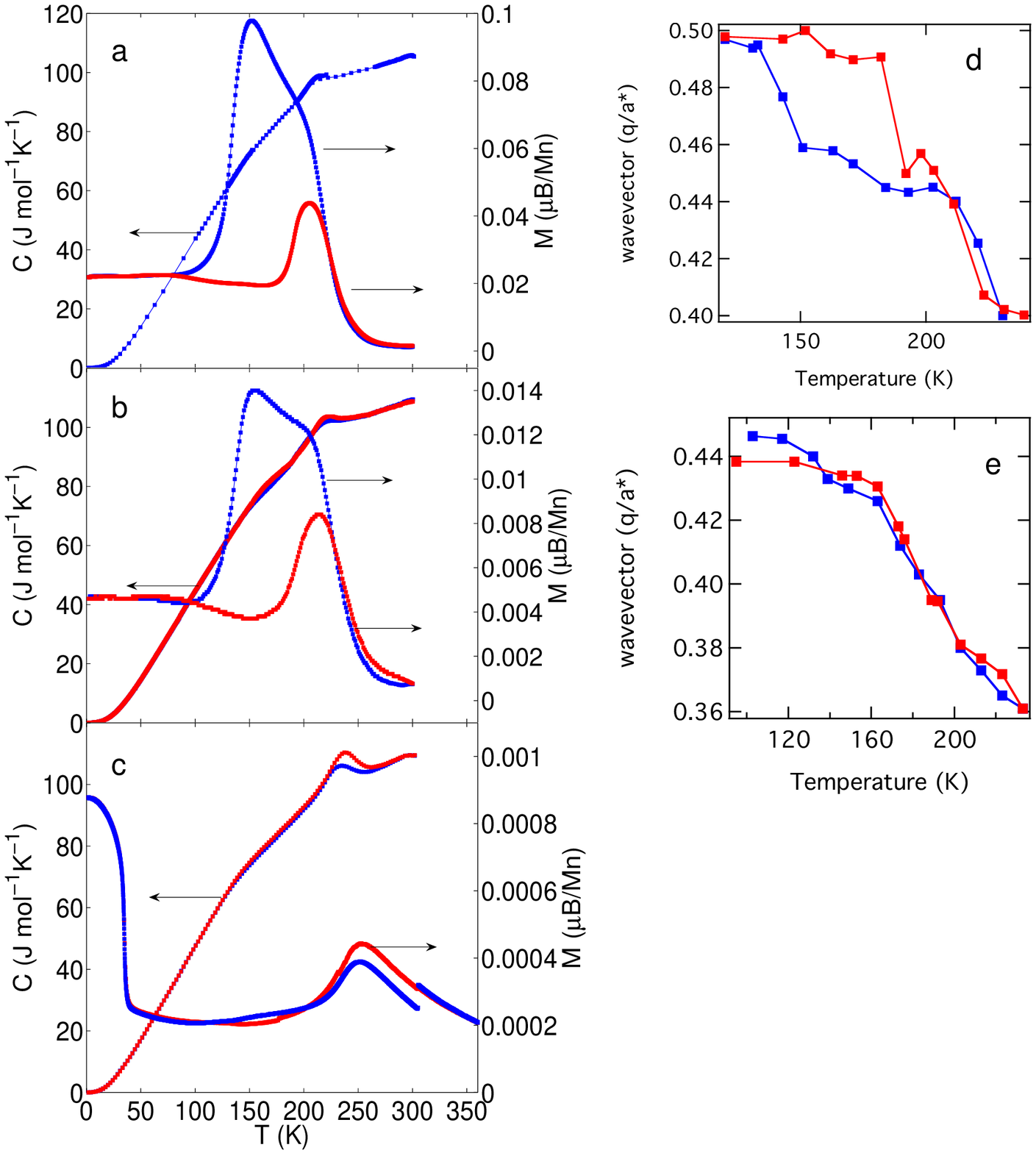} \caption{Heat capacity, 
and magnetisation (in 100 Oe),  with warming data shown in red and cooling data
shown in blue.  (a) shows data for \LCMO, (b) shows data for \LCMOfiftwo, and
(c) shows data for \PCMOfiftwo.  The errors are smaller than the size of the
data points.  The variation of the wavevector of the superstructure wavevector
with $T$ is shown for \LCMO in (d) (data from~\cite{chen_comm_incomm}) and for
\LCMOfiftwo in (e) (data from~\cite{us}).\label{lpcmo_ct_and_mt}}
\end{centering} \end{figure}

\begin{figure} \begin{centering}
\includegraphics[width=0.43\textwidth]{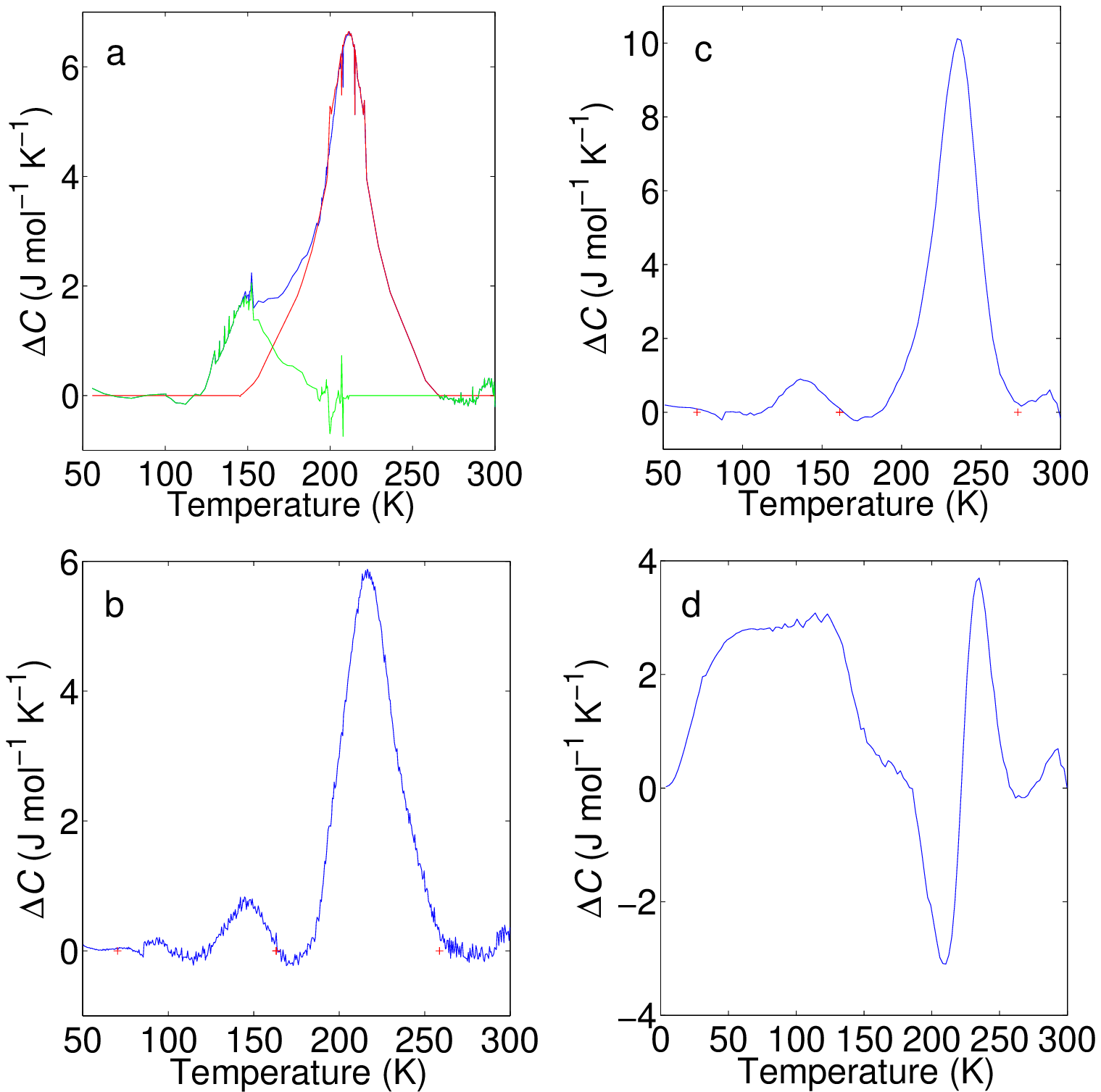} \caption{Excess heat
capacity of (a) \LCMO, (b) \LCMOfiftwo and (c) \PCMOfiftwo with background
removed.  (d) \LCMOfiftwo heat capacity subtracted from \PCMOfiftwo heat
capacity.  In (a) the blue line is total heat capacity above background. The
red and green lines show the contributions of the upper and lower $T$
transitions respectively.  In (b) and (c) the red crosses signify the limits
used for the entropy calculation.  \label{lpcmotrans}} \end{centering}
\end{figure}

The heat capacity data for all three compounds show two transitions (see
Fig.~\ref{lpcmo_ct_and_mt}), with the transition at higher $T$ exhibiting a
much larger change in entropy than the lower transition.  In order to make the
transitions more visible the background was removed from the data.  In the low
$T$ (1.8~K - 10~K) range, heat capacity data were fitted to an equation of the
form: \begin{equation} C_{\mathrm{P}} = \beta_3 T^3 + \beta_5 T^5 + \gamma T +
\frac{\alpha}{T^2} + \delta T^2 \ , \\ \end{equation} where $\beta_3$,
$\beta_5, \alpha$,  $\delta$ and $\gamma$ are constants, $\beta_3 = Nk \frac{12
\pi^4}{5} {\frac{1}{\theta_D^3}}$, where $\theta_D$ is the Debye
$T$~\cite{debye}.  The high $T$ data is  modelled with a Debye model and an
Einstein mode, where $\theta_\mathrm{D}$ has been determined from the low $T$
data.  The data were fitted with iterative reweighted least
squares~\cite{huber}  using the Levenberg-Marquardt method~\cite{marquardt}.
This method has the advantage over the more commonly used least-squares
technique that  low weights are automatically given to areas which have a poor
fit to the model (so the errors are not assumed to be Gaussian).  The heat
capacity above background is shown in Fig.~\ref{lpcmotrans}a,b, and c.

As can be seen from
Figs.~\ref{lpcmo_ct_and_mt}d,~\ref{lpcmo_ct_and_mt}e,~\ref{lpcmotrans}a and
~\ref{lpcmotrans}b the appearance of the superlattice reflections occurs at the
same $T$ as the upper transition, and the stabilisation of the value of the
wavevector occurs at the same $T$ as the lower transition.  This indicates that
the evolution of the superstructure is strongly linked to the phase
transitions.  The $T$s of the transitions (see Table~\ref{trans_temp}) for both
\LCMOfiftwo and \PCMOfiftwo show hysteresis. The lower transition shows greater
$T$ hysteresis than the upper transition, probably since at the lower $T$ the
superstructure has become more strongly pinned to grain boundaries and
impurities in the lattice. 

\begin{figure} \begin{centering}
\includegraphics[width=0.495\textwidth]{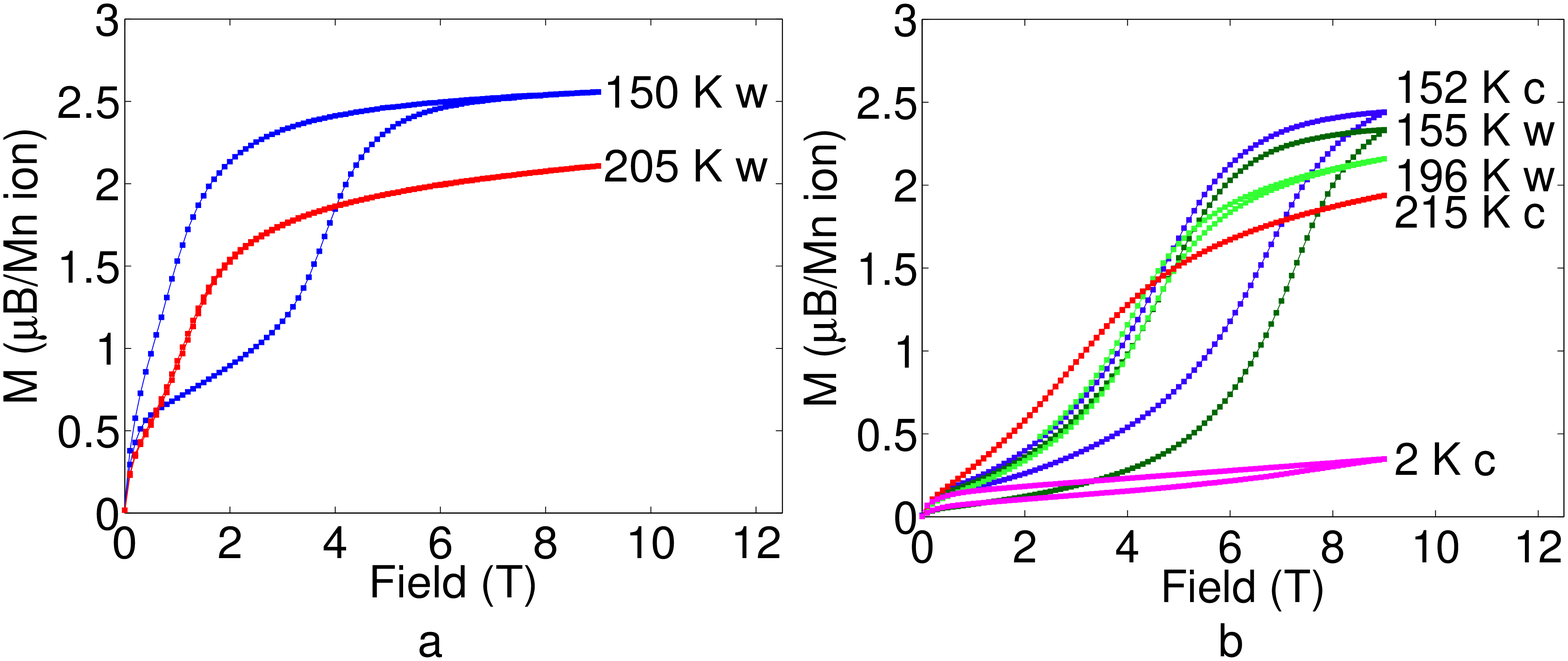} \caption{M-H loops taken at
various $T$s on warming and cooling for (a) \LCMO and (b) \LCMOfiftwo.  Curves
were taken on warming (w) and cooling (c).\label{lcmomh}} \end{centering}
\end{figure}

The magnetisation data for \LCMO and \LCMOfiftwo show an increase in magnetic
moment on cooling corresponding to a transition of some proportion of the
sample to FM (see Fig.~\ref{lpcmo_ct_and_mt}).  At lower $T$s the magnetisation
falls again - this is traditionally associated with the transition to
AFM~\cite{CMRoxidesb}.  In the $T$ range between the two transitions hysteresis
is observed in M-H loops (see Fig.~\ref{lcmomh}). 

The presence of an FM-AFM transition in \LCMO and \LCMOfiftwo agrees with the
suggestion of Milward \emph{et al.}~\cite{milward} that charge order which has
not locked into its low $T$ value will always be associated with
ferromagnetism.  The magnetisation is higher for \LCMO than for \LCMOfiftwo, as
predicted by Landau theory~\cite{milward}.

The magnetisation for \PCMOfiftwo shows a small change in the region of the
transitions.  However, the magnitude of the magnetisation for \PCMOfiftwo is
only 1\% of that for \LCMO.  However, there is also a marked increase in the
magnetic moment below 50~K  which is not associated with any obvious features
in the heat capacity data.  However, the \LCMOfiftwo and \PCMOfiftwo heat
capacity data do show a constant difference (on average 8\%) in the range 20 -
100~K (see Fig.~\ref{lpcmotrans}), with  an extra entropy of 5.1 J mol$^{-1}$
K$^{-1}$ arising in  \PCMOfiftwo relative to \LCMOfiftwo in this $T$ range.
Since the masses of Pr and La differ by only 1.4\%, the change in the phonon
contribution to the heat capacity is unlikely to have produced the large
observed difference~\cite{Loram}.  Thus there may be a magnetic phase in
\PCMOfiftwo which evolves continuously between 2 and 50~K.

The calculated entropies (see Table~\ref{entropy_values}) are  lower by a
factor of around two than those found in other studies of similar
compounds~\cite{Lees, Zheng}.  This difference is probably due to the fact that
the entropies in Refs.~\cite{Lees, Zheng} were calculated by fitting a
polynomial away from the region of the transitions, rather than using all of
the data to fit a background which is smooth in the region of the transitions,
as done in the present work. 

We now show that the two transitions are dominated by electron-lattice effects
associated with the superstructure.  At the upper transition, the percentages
of the total entropy are the same in \LCMO and \LCMOfiftwo (the absolute values
are only 28\% different), despite the fact that the magnetisation is ten times
smaller in \LCMOfiftwo.  In \PCMOfiftwo the entropy of the upper transition is
higher than that in \LCMO and \LCMOfiftwo , despite the fact that the
magnetisation is 100 times smaller in \PCMOfiftwo than in \LCMO.  Since
\PCMOfiftwo is expected to have stronger electron-phonon coupling than
\LCMOfiftwo, this indicates that the electron-phonon coupling is dominating the
value of the released entropy.  Thus the entropy of the upper transition is
dominated by electron-lattice effects, and shows little or no link with the
magnetisation.  The proportion of entropy released at the lower transition for
\PCMOfiftwo is only lower than the values for \LCMO and \LCMOfiftwo by a factor
of two, although the magnetic transition has all but disappeared.  Thus even
the lower transition is not dominated by magnetic effects.

When passing through a first order transition, the sample should emit latent
heat, producing a PPMS decay curve that cannot be well modelled by the analysis
software~\cite{PPMS}.  Therefore, decay curves were examined in the regions of
the transitions, but the values derived manually were the same as the values
which had been determined automatically, within experimental error.  This
absence of latent heat at the transition is the first piece of evidence that
the transitions are second order.

\begin{figure} \begin{centering}
\includegraphics[width=0.41\textwidth]{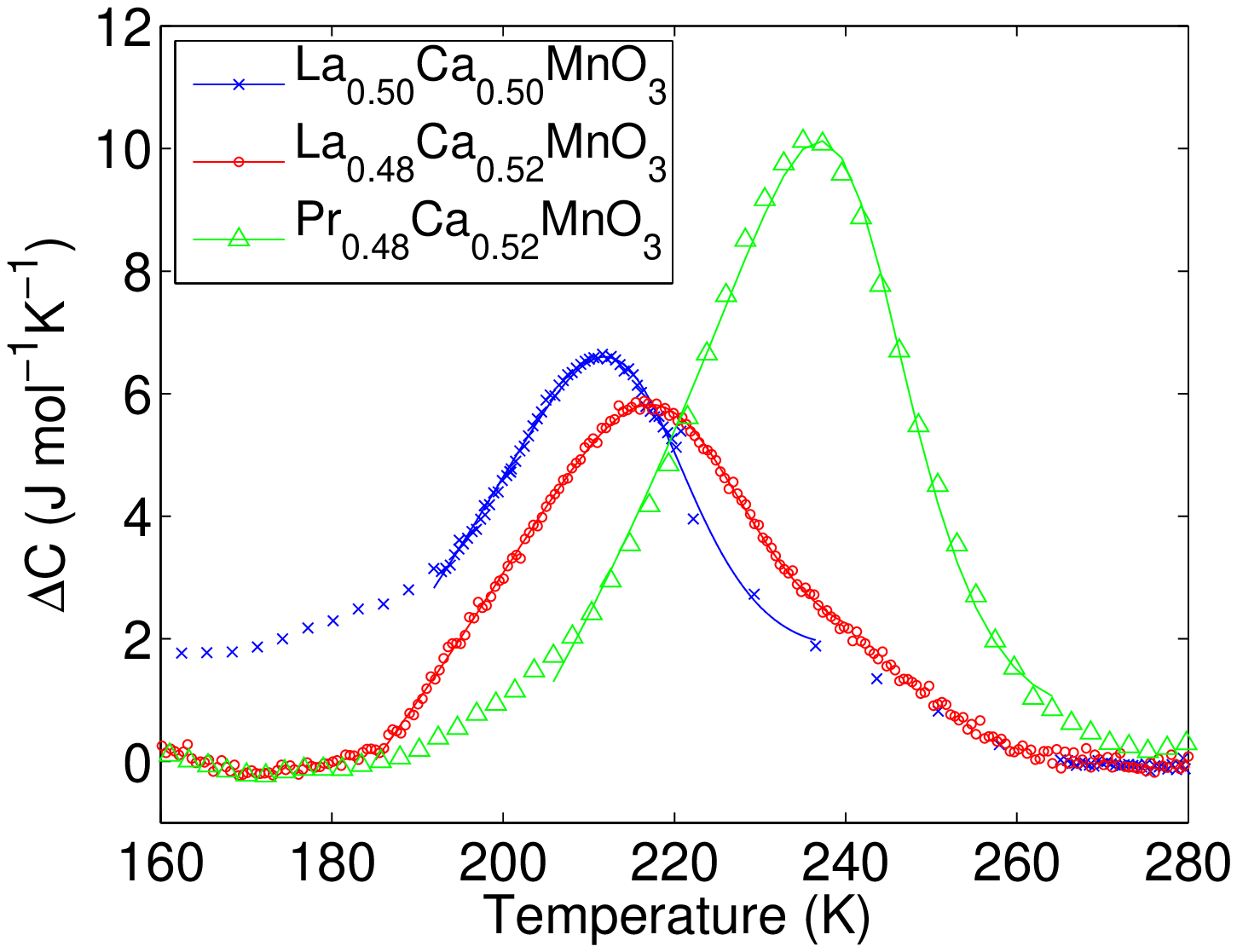} \caption{Fitting of
transition peaks for \LCMO, \LCMOfiftwo and \PCMOfiftwo to a model of a Peierls
transition in a system with impurities (a linear background was removed).
Points are data and the solid line is the fit of the model to the data.
\label{peakfit}} \end{centering} \end{figure}

The second piece of evidence that the upper transition is second order is that
the heat capacity peak is always asymmetric (see figure~\ref{lpcmotrans}).  A
second order transition in a very pure sample can be modelled using critical
exponents; however, the breadth of the peak in this case indicated that a model
including impurities must be used.  Therefore the heat capacity peak above
background at the transition was modelled as a Peierls transition in a system
containing impurities~\cite{Chandra}: \begin{equation} C \propto
\frac{d\chi}{dt} \propto \frac{d\kappa_\mathrm{imp}}{dt} \propto \frac{d}{dt}
\{(-t) + [(-t)^2 + N^4]^{1/2}\}^{1/2} \end{equation} where $\chi$ is the
magnetic susceptibility, $\kappa_\mathrm{imp}$ is the inverse ionic correlation
length in the presence of impurities, $N = \Lambda x^{1/d}$ and $t=
(T-T^*)/T^*$.  $\Lambda$ is an input lengthscale determined by the impurity
potential (taken to be roughly a lattice spacing), $d$ is the system dimension,
$x$ is the impurity concentration and
$T^*=T_\mathrm{C}^\mathrm{imp}=T_\mathrm{C}^\mathrm{pure} - \Delta T$.  Since
the low and high $T$ limits of this function are not the same, a linear
background was subtracted to enable the function to be fitted to the heat
capacity above background.  The fit can be made over the widest range for
\PCMOfiftwo because the upper transition is well separated from the lower
transition.  For \LCMO and \LCMOfiftwo the lower transition is close, and
therefore the fit must be made over a narrower range.  As can be seen from
Fig.~\ref{peakfit}, the model provides an extremely good fit to the data.

A lengthscale for the disorder was calculated as $x^{-1/d} = \Lambda / N$.  For
\PCMOfiftwo and \LCMO the result was 23~\AA, and for \LCMOfiftwo it was 21~\AA.
Therefore the lengthscale of the disorder is very similar in all three
compounds.  By comparison, blue bronze (a charge density wave system) doped
with 1\% W (K$_{0.3}$Mo$_{0.99}$W$_{0.01}$O$_3$) yields a lengthscale of 51\AA.
The fact that this model can fit all three compounds, with similar disorder
lengthscales, supports the conjecture that the transition is a Peierls
transition in disordered materials.  We suggest that the `impurities' may in
fact reflect the A-site cation inhomogeneity rather than chemical
inhomogeneities since x-ray and neutron data indicate that the samples are
single phase~\cite{Williams, pcmo_TEM}.

\begin{table} \begin{tabular}{|l|c|c|c|c|} \hline &  \multicolumn{2}{c|}{Lower
transition (K)} &    \multicolumn{2}{c|}{Upper transition (K)}\\ &  cool
&   warm            &  cool           		 & warm		\\ \hline \LCMO
&   		150		&		-     	&		223
&  -  \\ \LCMOfiftwo 	    &   		146		&
158  	&		218  	 &  223 \\ \PCMOfiftwo         &
131		&		138  	&  		229  	 & 234  \\
\hline \end{tabular} \caption{Transition $T$s for \LCMO, \LCMOfiftwo and
\PCMOfiftwo.\label{trans_temp}} \end{table}

\begin{table} \begin{tabular}{|l|c|c|c|c|} \hline & \multicolumn{2}{c|}{S of
lower transition}  &   \multicolumn{2}{c|}{S of upper transition}   \\ & J/(mol
K)  &  \% of S$_\mathrm{tot}$  & J/(mol K) & \% of S$_\mathrm{tot}$\\ \hline
\LCMO           &                   0.41           &        24               &
1.33    &    76    \\    La$_{0.48}$Ca$_{0.52}$MnO$_3$ &     0.25           &
21               &        0.95    &    79    \\ Pr$_{0.48}$Ca$_{0.52}$MnO$_3$ &
0.21              &        13               &        1.36    &    87    \\
Pr$_{0.6}$Ca$_{0.4}$MnO$_3$   &     0.6               &        23
&        2.0    &    77    \\  La$_{0.25}$Ca$_{0.75}$MnO$_3$ &     0.67
&        23                &        2.3    &    77    \\ \hline \end{tabular}
\caption{Entropy values for the transitions in various manganite compounds.
Data for Pr$_{0.6}$Ca$_{0.4}$MnO$_3$ taken from~\cite{Lees}, data for
La$_{0.25}$Ca$_{0.75}$MnO$_3$ taken from~\cite{Zheng}.\label{entropy_values}}
\end{table}

In conclusion, our data showed that the upper transition which has been
traditionally associated with the onset of FM in $x=0.5$ is in fact driven by
the lattice.  The transition is second order, and can be well modelled as a
Peierls transition in a disordered material.  The previous conclusion that the
transition was first order was  based merely on hysteresis in the resistivity
data~\cite{diaz2/3}, rather than on the measurement of any thermodynamic
quantity.   Such hysteresis can be explained as being due to lossy kinetics in
a CDW-like ground state with disorder~\cite{pcmo_TEM}.  Other work assumes this
transition is first order~\cite{Ramirez_cheong_schiffer} since the
electron-phonon coupling is taken to be large.  As we have shown, the
electron-lattice effects are dominant, but can be well modelled as a CDW in
which insulating behaviour can be produced without the need to invoke strong
electron-phonon coupling.  Finally, an unexpected low $T$ magnetic phase has
been found in \PCMOfiftwo which evolves continuously at low $T$.

We thank N.D. Mathur and N. Harrison for helpful comments.  Work at NHMFL is
performed under the auspices of the NSF, DoE and the State of Florida.  Work at
Cambridge was funded by the UK EPSRC and the Royal Society.  S. Cox
acknowledges support from the Seaborg Institute.

\end{document}